\documentclass[apj]{emulateapj}
\shorttitle{Red-sequence LF}
\shortauthors{Gilbank et al.}

\newcommand\gsim{\gtrsim}
\newcommand\lsim{\lesssim}
\newcommand\kms{km s$^{-1}$}
\newcommand\bgc{B$_{gcR}$}
\newcommand\zpri{$z^\prime$} 
\newcommand\rvir{r$_{\rm 200}$}
\newcommand\rzc{$(R-z^\prime)_{\rm corr}$} 
 
\begin{document}

\title{The Red-Sequence Luminosity Function in Galaxy Clusters since z$\sim$1.}

\author{David G. Gilbank\altaffilmark{*} and H. K. C. Yee}
\affil{Department of Astronomy and Astrophysics, University of Toronto,
  50 St George Street, Toronto, Ontario, M5S 3H4, Canada}
\email{gilbank@astro.utoronto.ca, hyee@astro.utoronto.ca}

\author{E. Ellingson}
\affil{Center for Astrophysics and Space Astronomy, University of Colorado at Boulder, CB389, Boulder, CO 80309, USA}
\email{Erica.Ellingson@colorado.edu}

\author{M. D. Gladders}
\affil{Department of Astronomy and Astrophysics, 
University of Chicago, 5640 S. Ellis Ave., Chicago, IL, 60637, USA}
\email{gladders@oddjob.uchicago.edu}

\author{Y.-S. Loh}
\affil{Department of Physics and Astronomy, University of California, Los Angeles, CA 90095-1547, USA}
\email{yeongloh@astro.ucla.edu}

\author{L. F. Barrientos}
\affil{Departamento de Astronom\'{\i}a y Astrof\'{\i}sica,
Universidad Cat\'{o}lica de Chile, Avenida Vicu\~na Mackenna 4860, Casilla 306, Santiago 22, Chile}
\email{barrientos@astro.puc.cl}

\and

\author{W. A. Barkhouse}
\affil{Department of Astronomy,
University of Illinois at Urbana-Champaign,
1002 West Green Street,
Urbana, IL 61801, USA}
\email{wbark@astro.uiuc.edu}

\altaffiltext{*}{Current address: Astrophysics and Gravitation Group, Department of Physics and Astronomy, University Of Waterloo, Waterloo, Ontario, N2L 3G1, Canada.  Email: dgilbank@astro.uwaterloo.ca}

\begin{abstract}
We use a statistical sample of $\sim$500 rich clusters taken from 72 square degrees of the Red-Sequence Cluster Survey (RCS-1) to study the evolution of $\sim$30,000 red-sequence galaxies in clusters over the redshift range 0.35$<$z$<$0.95.  We construct red-sequence luminosity functions (RSLFs) for a well-defined, homogeneously selected, richness limited sample.  
The RSLF at higher redshifts shows a deficit of faint red galaxies (to $M_V\ge$ -19.7) with their numbers increasing towards the present epoch.  This is consistent with the `down-sizing` picture in which star-formation ended at earlier times for the most massive (luminous) galaxies and more recently for less massive (fainter) galaxies.    We observe a richness dependence to the down-sizing effect in the sense that, at a given redshift, the drop-off of faint red galaxies is greater for poorer (less massive) clusters, suggesting that star-formation ended earlier for galaxies in more massive clusters.  The decrease in faint red-sequence galaxies is accompanied by an increase in faint blue galaxies, implying that the process responsible for this evolution of faint galaxies is the termination of star-formation, possibly with little or no need for merging.  At the bright end, we also see an increase in the number of blue galaxies with increasing redshift, suggesting that termination of star-formation in higher mass galaxies may also be an important formation mechanism for higher mass ellipticals.  By comparing with a low-redshift Abell Cluster sample, we find that the down-sizing trend seen within RCS-1 has continued to the local universe.

\end{abstract}

\keywords{
galaxies: clusters: general -- galaxies: elliptical and lenticular, cD -- galaxies: evolution -- galaxies: luminosity function, mass function
}

\section{Introduction}
Clusters of galaxies are ideal laboratories for studying galaxy evolution since they contain many galaxies seen at the same epoch in close proximity.  Their cores are dominated by early-type galaxies, which are the major component of the high mass end of the galaxy stellar mass function locally.  There is now a good deal of evidence that cluster early-type galaxies formed the bulk of their stars at high redshift and thereafter simply evolved passively with little or no residual star-formation.  One such line of evidence is the tight sequence they form in color--magnitude space  \citep[e.g.,][]{visv,bow92}, `the red-sequence'.  A similar red-sequence is also seen for early-type galaxies in the field out to at least z$\sim$1 \citep{Bell:2004lb}. Furthermore, {\it all} galaxies appear to be divided into two distinct populations: the passively-evolving red-sequence and the actively star-forming `blue cloud'.  Only a small amount of residual star-formation (less than $\sim$10\% of the galaxies' past averaged star-formation rate) is necessary to move a galaxy from the red-sequence to the blue cloud.  Therefore, early-type galaxies can provide unique insight into the history of star-formation, as traced by objects in which star-formation has already been terminated.

In the local universe, the probability of a galaxy belonging to the red-sequence or blue cloud depends on its stellar mass and its environment \citep{Baldry:2006bq}.  It is likely that the other fundamental parameter governing the properties of a galaxy is the epoch at which it is observed.  Thus, in order to build a complete picture of galaxy evolution, we need to study the colors of galaxies as a function of mass (or luminosity), environment and redshifts.

The classical picture for the formation of galaxies proposes a single `monolithic collapse` \citep{Eggen:1962yu}, with stars in elliptical galaxies being formed in a single burst, thereafter evolving passively \citep{Partridge:1967nh,Sandage:1970lj}.  This very simple model predicts remarkably well many of the properties and scaling relations of elliptical galaxies.

In the current hierarchical paradigm, structure forms in a `bottom-up' sense, as galaxies and clusters are built from the merging of smaller units.  Recently, there is growing evidence that star-formation has evolved in a `top-down` sense with more massive galaxies being most actively star-forming in the past and the bulk of the star-formation activity moving toward less massive galaxies  as the universe ages.  Although this seems intuitively at odds with hierarchical models, scenarios have been proposed in which star-formation progresses in this anti-hierarchical manner \citep{De-Lucia:2006sa}.  Whereas previous generations of semi-analytic models in this hierarchical framework suggested that the most massive early-type galaxies should be younger than less massive ones \citep{Baugh:1996vs,Kauffmann:1998mh}, in order to reconcile with the observed down-sizing trend, the prediction is now that although the most massive early-types assembled their {\it mass} later than lower mass early-types, the stellar mass has been built up through a series of gas-poor mergers which do not result in additional star-formation. Hence the earlier formation times of the stellar populations in more massive galaxies is recovered.

Despite numerous signs of merging in early-type galaxies \citep[e.g.,][]{van-Dokkum:2005dl,Tran:2005xc}, it remains an open question how important mergers are in their formation and evolution.  The problem of disentangling how a galaxy assembled its mass from how it assembled its stars is a difficult one.  

Several studies of field galaxies have reported this down-sizing or anti-hieararchical trend in star-formation \citep[e.g.,][]{Bell:2004lb,Juneau:2005ft,Faber:2005yw,Bundy:2005bc,Scarlata:2007oz}.  Initial results suggested that the comoving number density of massive early-type galaxies had evolved more than could be accounted for by passive evolution alone, and that `dry merging` of massive galaxies was required \citep{Bell:2004lb,Faber:2005yw}.  More recently, it has been suggested that most, if not all, of the evolution can be attributed to the termination of star formation and pure passive evolution; and that a significant contribution from merging is not required \citep{Cimatti:2006vz,Scarlata:2007oz}. 

In galaxy clusters, down-sizing appears to be supported by the spectroscopic ages of red-sequence galaxies as a function of mass \citep{Nelan:2005ml}.  In distant clusters (z$\sim$0.8), a deficit of faint red-sequence galaxies relative to local clusters has been claimed, in accordance with this picture \citep{De-Lucia:2004xa,Tanaka:2005mk,de-Lucia:2007li}.  However, all of these high-redshift works have found a large cluster-to-cluster scatter in their samples of sizes of approximately 1-10 clusters, indicating that a large, statistical sample is crucial to such studies.  The relative contributions of passive evolution versus dry merging to explain the evolution of the number density of early-type galaxies both in the field and in clusters is still an open question.

In this paper we present results using the first statistical sample of galaxy clusters drawn from a well-defined, wide area, homogenous survey covering a large redshift range, 0.35$<$z$<$0.95.  We present the survey data in \S2 and detail our method for constructing composite clusters in \S3.  In \S4 we examine the Red-Sequence Luminosity Function and use the ratio of luminous-to-faint red-sequence cluster galaxies to trace its evolution with redshift and dependence on cluster mass. In \S5 we discuss our results and compare with other studies of early-type galaxy evolution both in clusters and the field, and in \S6 we present our conclusions.  Throughout we assume a cosmology of $H_0=$70 km s$^{-1}$Mpc$^{-1}$ (and $h=H_0/$100 km s$^{-1}$Mpc$^{-1}$), $\Omega_{\rm M}=$0.3 and $\Omega_\Lambda=$0.7.

\section{Data}
The Red-Sequence Cluster Survey \citep[RCS-1][]{Gladders:2005oi} is a two filter imaging survey covering $\sim$100 square degrees.  It was designed to build a well-defined sample of galaxy clusters out to z$\sim$1 using a highly-efficient color selection technique \citep{gy00,Gladders:2005oi}.  This technique provides a photometric estimate of the cluster redshift, accurate to $\Delta$z$\sim$0.05 (e.g., \citealt{blindert07a,gilbank:07a}).  A proxy for the mass of each cluster is produced by measuring the optical richness, which is obtained by calculating the amplitude of the galaxy-cluster cross-correlation function, $B_{\rm gc}$ \citep{ls79,ylc99}.  We use a modified version of the B$_{\rm gc}$ parameter \citep[\bgc, see][]{Gladders:2005oi}, considering only galaxies with colors compatible with the red-sequence at the estimated redshift of the cluster.  Although the uncertainty on \bgc~for an individual cluster is relatively large ($\sim$20-30\%), the accuracy for ensembles is good, as demonstrated by the agreement between cosmological parameters derived from \bgc-selected samples of RCS-1 clusters and the current best-fit cosmology from other methods \citep{Gladders:2007us}.  Extensive work is on-going to calibrate the scaling relation between \bgc~and mass from RCS clusters.  Some early results have been presented in \citet{blindert07a, gilbank:07a, felipe07}.  Results from the X-ray selected CNOC1 clusters can be found in \citet{Yee:2003we,Hicks:2006ap}.

We use data from the RCS-1 photometric catalogs, which are derived from moderate depth imaging data in the $R_{\rm C}$- and \zpri-bands.  The imaging was obtained with two mosaic cameras on 4-m class telescopes, {\it CFH12K} on the CFHT and {\it MOSAIC-II} on the CTIO Blanco telescope.  Details of the data reduction are given in \citet{Gladders:2005oi} and we only give a brief account here.  The survey is divided into 20 patches, each typically around 2 degrees $\times$ 2 degrees.  Object detection, classification and photometry were performed on the images using {\sc PPP} \citep{1991PASP..103..396Y}. For each object, total magnitudes in the deeper of the two filters (usually $R$) were computed from a curve of growth analysis.  Colors were measured using a 3\arcsec~aperture or the optimal aperture from the curve of growth, if it is smaller.  Total magnitudes for the shallower filter were then calculated using the total magnitude of the deeper filter and this color.  Galactic extinction was corrected using the maps of \cite{1998ApJ...500..525S}.  The \zpri~magnitudes are expected to be uncertain at the \zpri$\lsim$0.10 level, and this is confirmed by our own comparison of internal overlaps (and also additional photometry from other follow-up imaging) and the $(R-z^\prime)$ colors should be accurate to $\lsim$0.03 \citep{Gladders:2005oi}.  Throughout, we use magnitudes on the AB system unless otherwise specified.  In this paper we only consider the 72 square degrees selected to have the highest photometric quality (see \citealt{Gladders:2007us} for details).  

\section{Constructing composite clusters}
\label{sec:comp}
We construct composite clusters following the basic technique of \citet{Loh:2007be}.  We repeat the salient points of that work here, and add details pertinent to our analysis.  To build our sample, we select clusters from the latest (December 2005) generation RCS-1 cluster catalog, with red-sequence estimated redshifts 0.35$<$ z$_{RCS}\le$0.95; richness, \bgc$>$500 (in units of $(h_{50}^{-1}$ Mpc$)^{1.77}$); and detection significance, $\sigma_{\rm RCS}\ge$3.3-$\sigma$. A \bgc$=$500 cluster corresponds to a velocity dispersion $\sigma \sim$400\kms~\citep[e.g.,][]{blindert07a}, and our sample extends up to \bgc$\sim$2000 clusters which would correspond to $\sigma \sim$1200\kms.

For each cluster, we extract colors and magnitudes (binning 0.1 magnitudes in \zpri-band magnitude and 0.05 magnitudes in $(R-z^\prime)$ color) for all galaxies within a radius of 0.5$\times$\rvir, where the value of \rvir~is estimated from the richness following the relation $\log r_{\rm 200}=0.48 \log B_{\rm gc} -0.95$ (\citealt{barkhouse07a}, after converting to our cosmology from their $h=0.5$ to $h=0.7$; see also \citealt{Yee:2003we}).  This corresponds to a physical radius of between $\sim$1.1 $h^{-1}$Mpc and 1.9$h^{-1}$Mpc for the range of richnesses used here.  For some clusters, the circle defined by this radius may fall partially off the edge of our survey.  We use detailed maps of the positions of all the CCD chips to calculate the fractional area lost due to survey geometry.  We use this fraction to reject clusters where the fractional areal completeness is $<$0.6 and to correct incomplete clusters.  The total fraction of clusters rejected due to geometric considerations alone is around 20\%.  The rejected clusters are not a strong function of redshift or richness, so the net effect is simply to reduce the total usable area of the survey by $\sim$20\%.  For the remaining clusters, more than 80\% of the sample have fractional areal completenesses of $>$0.8.

The data from these clusters are placed into redshift bins of $\Delta$z$=$0.1.  The typical uncertainty in redshift given the red-sequence color of a cluster varies from $\sim$0.04 to 0.08 within RCS-1 (e.g., \citealt{blindert07a}, \citealt{gilbank:07a}).  We do not attempt to correct the colors of each cluster within each redshift bin to a common redshift, since the systematic difference in color is smaller than the size of the random error due to the accuracy of measuring the position of the red-sequence in color-magnitude space, and doing so will in fact increase the dispersion of the composite color magnitude diagrams (CMDs). 

The composite CMDs so created contain cluster galaxies plus contamination from background/foreground galaxies.  We remove this contamination in a statistical way by creating background CMDs by summing the data from all galaxies within RCS-1.  In practice, we sum galaxies within each RCS-1 patch (disjoint areas of sky of typically $\sim$4 deg$^2$) and create a composite background by summing the background from each patch, weighted by the number of clusters each patch contributes to our final cluster sample.  We do not explicitly remove clusters from these background fields, as the total area contributed by all the clusters within any patch is small ($\sim$2\%) and masking the clusters each time is computationally expensive.  We verified that performing this masking does not affect our final results.

Errors are propagated through for each bin of the CMD accounting for the Poisson uncertainty due to cluster counts, background counts, and the variance from patch-to-patch.  The number of cluster galaxies in the $j^{th}$ bin of the composite cluster CMD, $N_{cj}$, is given by:

\begin{equation}
N_{cj} = \sum_i N_{ij} - n_{fj},
\end{equation}
where $N_{ij}$ is the observed number of galaxies (cluster $+$ field) in the $j^{th}$ bin of the $i^{th}$ cluster region ($N_{cj}$ is allowed to be formally negative in our method. In practise we shall be binning over many bins of color and magnitude in our analysis, such that the totals in these larger bins are always positive).  The field contribution, $n_{fj}$ is given by: 

\begin{equation}
n_{fj} = \frac{\sum_i \Omega_i}{\sum_p \Omega_p} \sum_p N_{pj} w_p,
\end{equation}
where $\Omega_i$ and $\Omega_p$ represent the areas of the cluster regions and the $p^{th}$ patch (field) respectively.  The patch weighting is:

\begin{equation}
w_p = \frac{m_p}{\sum_p m_p},
\end{equation}
where $m_p$ is the total number of clusters in patch $p$ contributing to the composite cluster. This gives more weight to data from those patches which contribute more clusters, usually due to the greater area they cover or greater uniformity of data.

The error in the counts of the $j^{th}$ bin is the quadrature sum of cluster and field contributions
\begin{equation}
\sigma (N_{cj}) = [\sigma (n_{ij})^2 + \sigma (n^\prime_{fj})^2]^{1/2}.
\end{equation}
We denote the field uncertainty here as $n^\prime_{fj}$ since we allow for the Poisson uncertainty, $n^{1/2}_{fj}$, plus the patch-to-patch standard deviation, $\sigma^p_{fj}$, in a manner akin to that of \citet{edcc}

\begin{equation}
\sigma (n^\prime_{fj}) = max(n^{1/2}_{fj}, \sigma^p_{fj}).
\end{equation}
The rationale for a semi-global background subtraction (i.e., the approach of Eq.~1-5) is discussed in more detail in \citet{Loh:2007be}.

Once a background subtracted cluster CMD has been created, the next step is to fit the red-sequence.  Several techniques were tried to fit the locus in the presence of the ``blue cloud" galaxies.  The method traditionally used is to fit using the biweight estimator \citep{tukey}.  We found that using this method on the whole CMD (particularly in the highest redshift bins) caused poor fits, mainly due to the significant population of blue galaxies at faint magnitudes which biased the fit, causing the relation to be much steeper than that estimated by eye.  To circumvent this problem, we overlaid model tracks for the expected red-sequence and rejected galaxies bluer than 0.2 magnitudes in $(R - $\zpri$)$ than the expected colors.  This effectively isolates the red-sequence (as confirmed by visual inspection), but still allows the slope and intercept of the relation to be fine-tuned.  We also found that reducing the scale radius used to select red-sequence galaxies to 0.25$\times$\rvir~allows a cleaner fit and so we use this value for fitting the CMR.  The results of the fit are given in Table~\ref{tab:cmrfit}.

Next, we subtract the best-fit color-magnitude relation to leave a red-sequence which is horizontal in color-magnitude space, resampling the binned CMD using sinc interpolation.  The result is a red-sequence centered on zero color, with no magnitude dependence.  We denote these corrected colors as \rzc. Blue galaxies will have a stronger k-correction than red galaxies, and so we apply a differential k-correction to the blue galaxies, as a function of their observed color (see e.g., \citealt{Loh:2007be}). The primary effect of this correction is to dim the contribution of on-going star formation in the brightest blue galaxies, allowing cleaner selection of the brightest red-sequence galaxies. With these corrections, the resulting magnitude of each galaxy is thus now more closely related to its stellar mass.  This correction is discussed in more detail in \S\ref{sec:totlf}.

\begin{table}
\caption{Fitted CMR parameters for the composite clusters. \label{tab:cmrfit}}
\centering{
\begin{tabular}{lccc}
\tableline\tableline
$\bar{z}$ & $m^\star_{z^\prime}$ & $(R-z^\prime)$ at  $m^\star_{z^\prime}$ & d$(R-z^\prime)$/d\zpri\\
\tableline
  0.40 & 18.940 &  0.825  & -0.044\\
  0.50 & 19.453  & 0.925  & -0.046\\
  0.60 & 19.908  & 1.063  & -0.048\\
  0.70 &  20.322  & 1.243 & -0.049\\
  0.80 &  20.675  & 1.439  &-0.059\\
  0.90 &  20.955  & 1.632 & -0.063\\
\tableline
\end{tabular}
}
\tablecomments{Columns show the average redshift of the composite cluster, the observed value of $m^\star$ in \zpri~(derived from the z$=$0.4 fit and the passively-evolving model), the observed color at this magnitude and the slope of the relation.}
\end{table}

Finally, to isolate red-sequence galaxies for further study, we choose to use only galaxies on the red side of the red-sequence, i.e., \rzc$\ge$0.  This eliminates contamination by galaxies blueward of the red-sequence whose magnitude errors may allow them to scatter onto the red-sequence.  Since red-sequence galaxies are the reddest normal galaxies at a given redshift, there should be no galaxies redward of them, after background subtraction.  We verified the CMR goodness-of-fit by centroiding the CMD data in color about the \rzc$=$0 line and applying a small sub-pixel shift, if necessary.  Such color shifts were $\lsim$0.01.  The red-sequence was then extracted by mirroring this distribution about the \rzc$=$0 line.  To include the effects of brightest cluster galaxies (BCGs) which are often found to be slightly bluer than the red-sequence, possibly due to the effects of on-going star formation due to the accretion of cold gas at the cluster center \citep[e.g.,][]{McNamara:1992wj}, we relax the requirement of \rzc$\ge$0 to \rzc$\ge$-0.5 for galaxies brighter than $M^\star$.  This allows the inclusion of galaxies which are clearly seen to be separate from the blue galaxy population (after application of the differential k-correction, described above).

\section{Red-sequence luminosity functions}
\label{sec:rslf}
\begin{figure}
\vspace*{-5mm}
\plotone{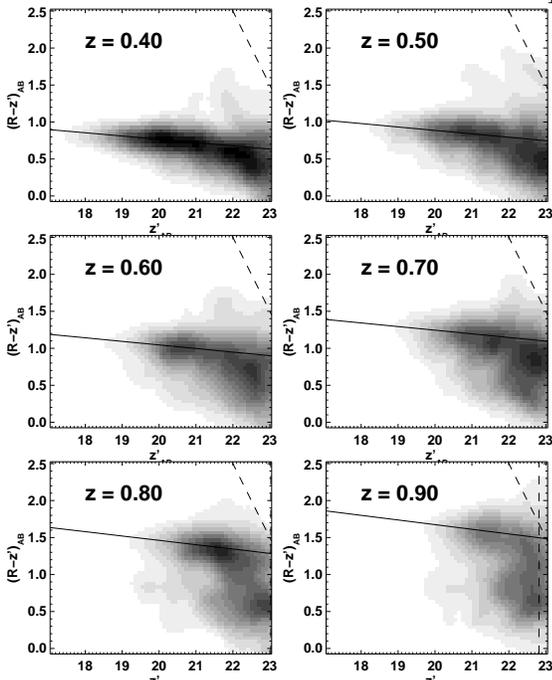}
\caption{Color-magnitude diagrams for the background-subtracted composite clusters in redshift bins of 0.1. Solid line indicates fit to red-sequence (Table 1). Dashed lines represent the 100\% completeness limits adopted. Note that while the completeness in \zpri~of the RCS catalog is somewhat deeper, we adopt the limits marked by the dashed lines so that the catalog is complete in $(R-z^\prime)$ color to colors significantly redder then the red sequence (see text).  The 2D histograms have been Gaussian smoothed for display purposes. 
}
\label{fig:cmds}
\end{figure}

We are now left with a CMD, constructed to contain only red-sequence galaxies.  Red-sequence luminosity functions can be constructed by simply summing over the color bins.  A crucial step is to understand the magnitude completeness of the sample.  Modelling the incompleteness, as is usually done for galaxy number counts in the field, becomes a much more complicated problem due to the color cuts imposed.  The only reliable way to verify the completeness for red-sequence selected galaxies is by deeper imaging of the same areas of sky \citep[e.g.,][]{Cimatti:2006vz}.  We adopt a very conservative approach and cut our data to a magnitude limit which should provide close to 100\% completeness for galaxies \citep{1991PASP..103..396Y}.  We adopt a limit 0.8 magnitudes brighter than the 5-$\sigma$ point source magnitude limits \citep{Gladders:2005oi}.  This produces color and magnitude limits in observed $R$ and \zpri.  To ensure that we are not incomplete for the faintest, reddest galaxies, we calculate the intercept of the $R$-band magnitude limit with the red envelope of the red-sequence, prior to removing the red-sequence slope, and decrease the \zpri~mag limit to this value (typically $\approx$0.2 mags brighter).  This offset can been seen in Fig.~\ref{fig:cmds}.  

Since we only need the bluer ($R_{\rm C}$-band) data to measure the color of each galaxy, we can relax the limit from 0.8 magnitudes brighter than the 5-$\sigma$ point source limit ($M_{5\sigma, \rm R} - 0.8$) to only $M_{5\sigma, \rm R} - 0.3$.  The typical color errors show that we are still measuring colors of the faintest galaxies with an uncertainty of $\lsim$0.15 magnitudes at the faintest limit adopted.  We note that repeating our analysis with the very conservative magnitude limits of $M_{5\sigma} - 0.8$ in both filters does not change any of our results at all, except to give us insufficient depth to make reliable measurements in our highest redshift bin (0.85$<$z$\le$0.95), described later.

\begin{figure*}
\epsscale{1.}
\plotone{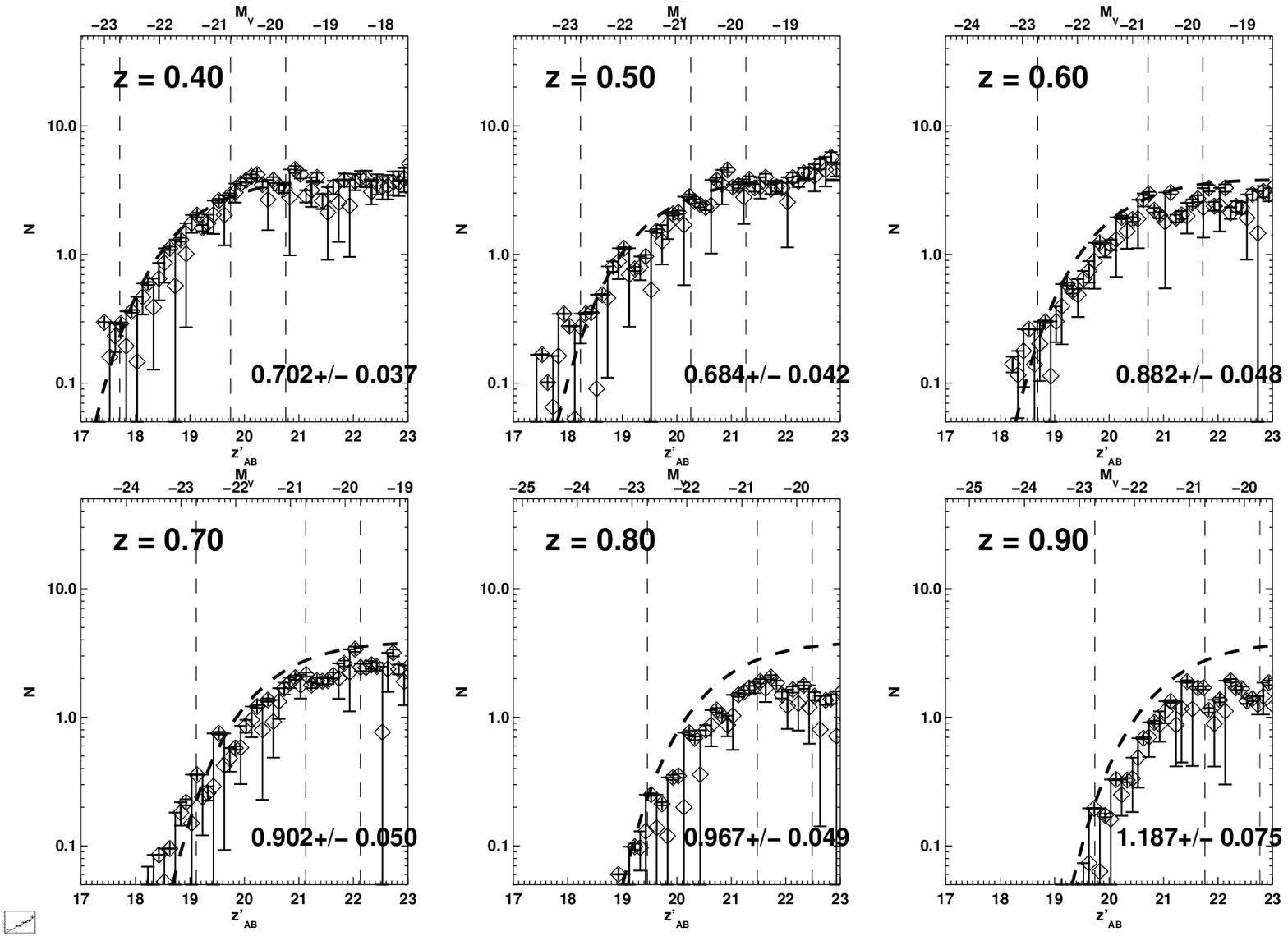}
\caption{Luminosity functions for red-sequence galaxies in the composite clusters shown in Fig.~\ref{fig:cmds}.  The mean redshift of each composite cluster is given in the upper left of each panel. Data are only shown down to the 100\% completeness limit for galaxies.  Error bars are dominated by uncertainty due to background subtraction (as discussed in \S\ref{sec:comp}).  The thick dashed line is the best-fit Schechter function of the lowest redshift composite cluster, evolved according to passive evolution to each redshift.  The upper x-axis shows the rest-frame absolute V-band magnitude and vertical dashed lines indicate limits of bright and faint bins adopted in our analysis. 
\label{fig:lfs}
}
\end{figure*}

Fig.~\ref{fig:lfs} shows the red-sequence luminosity functions.  We remind the reader that we have constructed these by averaging over all clusters of richness \bgc$>$500 within 0.5$\times$\rvir.

\citet{colless89} constructed composite luminosity functions by using:

\begin{equation}
N_{cj} = \frac{N_{c0}}{m_j} \sum_i\frac{N_{ij}}{N_{i0}},
\end{equation}
where $N_{cj}$ is the number of galaxies in the $j$th bin of the composite LF, $N_{ij}$ is the number in the $j$th bin of the $i$th cluster's LF, $N_{i0}$ is the normalization of the $i$th cluster LF, $m_j$ is the number of clusters contributing to the $j$th bin and $N_{c0} = \sum_i N_{i0}$.  This method is optimized for finding the composite LF under the assumption that it is universal (so that a simple rescaling by the cluster richness is all that is needed to find the average cluster LF), and in the presence of cluster data extending to different depths for different clusters.  The approach we adopt here is effectively setting $N_{i0}$ to unity, as we are trying to examine the average luminosity function of all clusters within our uniformly selected sample of clusters, at each redshift.  By integrating over a well sampled volume of the universe, the weighting intended by Colless' $N_{i0}$ term actually occurs naturally, since less rich clusters (containing fewer galaxies) are more abundant and hence their contribution is 'up-weighted' relative to richer clusters.  The actual number of clusters and the red-sequence galaxies they contribute are shown for various samples in Table \ref{tab:nums}.

We convert our observed \zpri-band magnitudes to rest frame $V$-band, using the same method as \citet{de-Lucia:2007li}.  We use the {\sc GALAXEV} stellar population synthesis code \citep{Bruzual:2003de} to generate model galaxy SEDs arising from a single burst stellar population formed at z$_f=$3.  We use three populations of different metallicities.  The zeropoint of each is normalized such that the observed CMR of Coma \citep{Terlevich:2001bx} is reproduced.  These also give reasonable agreement with our observed CMRs as a function of redshift.  Observed magnitudes are converted to rest frame magnitudes, interpolating between the nearest models.  We choose not to explicitly fit Schechter functions \citep{schechter} to each redshift bin as a) the characteristic magnitude, $M^\star$, and faint end slope, $\alpha$, are degenerate; and b) in the higher redshift bins, our data are not sufficiently deep to place strong constraints on the faint end slope using this parametric fit.  Instead, to study the evolution of the number density of red-sequence galaxies, we fit a Schechter function to the lowest redshift bin and passively evolve it using the above model to other redshifts.  The data in the lowest redshift, 0.35$<$z$\le$0.45, bin are sufficiently deep that a single Schechter function is not an adequate fit for the very faintest galaxies and it can be seen that the well-known \citep[e.g.,][]{barkhouse07a} upturn occurs for dwarf galaxies at $M_V\gsim$-18.  However, this fit is sufficient to be illustrative at brighter magnitudes.  For reference, the fitted parameters for this reference model are $\alpha$=-0.94$\pm$0.04 and $m_{z^\prime}^\star=$18.94$\pm$0.09, which corresponds to an absolute $M_{z^\prime}^\star=$-21.51.

\begin{table}
\caption{Number of clusters and red-sequence galaxies in each redshift bin for the different samples. \label{tab:nums}}
\begin{tabular}{lcccccc}
\tableline\tableline
$\bar{z}$ & \multicolumn{2}{c}{\bgc$>$500} &  \multicolumn{2}{c}{300$<$\bgc$\le$500} & \multicolumn{2}{c}{\bgc$>$800} \\ 
 & \multicolumn{2}{c}{main sample} & \multicolumn{2}{c}{`poorer' sample} & \multicolumn{2}{c}{`richer' sample} \\
 & $N_{\rm clus}$ & $N_{\rm gal}$ & $N_{\rm clus}$ & $N_{\rm gal}$ & $N_{\rm clus}$ & $N_{\rm gal}$ \\
\tableline
0.40 & 57  & 8378 & 54 & 4981 & 22 & 4463\\ 
0.50 & 52  & 7234 & 75 & 3942 & 9 & 1840 \\
0.60 & 70   & 5901 & 84 & 3704 & 16 & 1682\\ 
0.70 & 80   & 5765 & 82 & 3194 & 20 & 2036 \\
0.80 & 103  & 4253 & 104 & 2625 & 27 & 1299\\ 
0.90 & 98   & 3883& 63 & 1165 & 29 & 1889 \\
\tableline
\end{tabular}
\tablecomments{$N_{clus}$ refers to the number of clusters used for the analysis, after rejecting those with too low areal completeness, etc.;  $N_{gal}$ refers to the number of red-sequence galaxies down to the 100\% completeness limit.}
\end{table}

\subsection{Luminous-to-faint ratios}

\citet{Tanaka:2005mk} and \citet{de-Lucia:2007li} both used the ratio of luminous-to-faint galaxies to look for evolution in the faint end of the red-sequence luminosity function with redshift.  Our data are not as deep as either of these two works.  However, we have sufficient depth to use the \citet{de-Lucia:2007li} magnitude limits ($M_V \ge -20$ and $-20 < M_V \le -18.2$) to z$=$0.5, so first we compare our luminous-to-faint ratio with theirs.  For their z$=$0.5 composite EDisCS cluster, \citet{de-Lucia:2007li} find a luminous-to-faint ratio of 0.598$\pm$0.064 (for their sample utilizing photo-z selected plus statistical substraction) or 0.695$\pm$0.077 (for statistical subtraction only), measured in the observed $I$-band.  We find a value of 0.644$\pm$0.029 measured in the observed \zpri-band.  Our data probably more closely resemble their latter method, since we do not use photometric redshifts to reject galaxies on an individual basis.  Our measurement agrees with their value to within the uncertainty.

\begin{figure}
\plotone{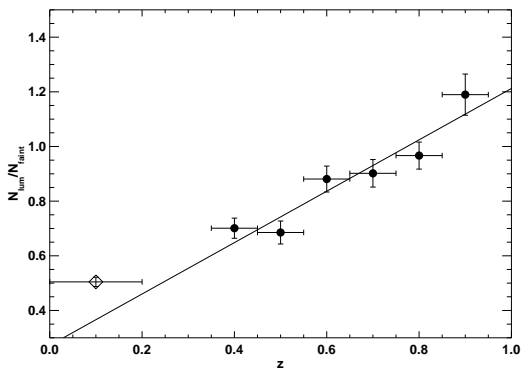}
\caption{Evolution of the ratio of luminous-to-faint red-sequence galaxies with redshift.  Horizontal error bars represent redshift range used in each bin.  Line is best linear fit accounting for errors.  The open diamond shows a low-redshift comparison point (not included in the fit) built from \citet{barkhouse07a} data, discussed in \S\ref{sec:lowz}.
}
\label{fig:evol_gd}
\end{figure}

In order to reach a reasonable redshift to search for evolution in the LF, say z$\sim$0.8, we must adopt a brighter magnitude limit.  We choose to define luminous galaxies as -22.7$< M_V \le$-20.7 and faint as -20.7$ < M_V \le$ -19.7.  These limits allow us to use data up to our 100\% completeness limit at z$=$0.9, avoid including BCGs at the bright end, and provide approximately equal numbers of galaxies in each luminosity bin, at z$=$0.9.  These bounds are shown as dashed lines in Fig.~\ref{fig:lfs}.  The number in the lower right of each panel denotes the value of this ratio and its error.  We calculate this value and its error following the same method as described in \S3 and this can be thought of as a limiting case where $N_{cj}$ reduces to a luminosity function of two magnitude bins.  We show the evolution of the luminous-to-faint ratio in Fig.~\ref{fig:evol_gd}.  A simple linear fit with redshift is sufficient, $\propto z^\beta$ with $\beta=$0.94$\pm$0.18. 

This trend appears to be robust to changes in the exact choice of magnitude limits for the two bins, provided that the breakpoint is chosen to be $M_V \sim -$21 or fainter.

\subsection{Cluster mass dependence of luminosity function}

\begin{figure}
\plotone{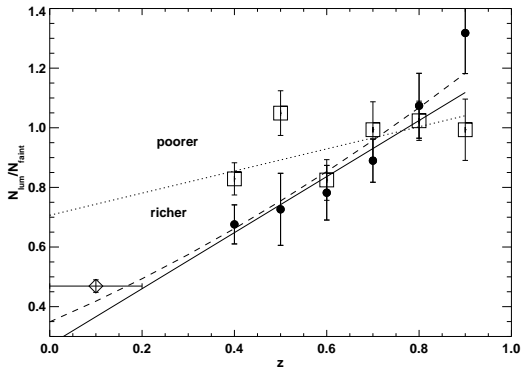}
\caption{As Fig.~\ref{fig:evol_gd}, but splitting the cluster sample into richer (\bgc$>$800, filled circles) and poorer (300$<$\bgc$\le$500, open squares) bins.  Solid line is the best fit from Fig.~\ref{fig:evol_gd} showing the fit to the whole sample, and the dotted line is a linear fit to the poorer clusters.  Dashed line shows a $(1+z)^\beta$ fit to the rich clusters including a low redshift rich composite cluster based on \citet{barkhouse07a} data, discussed in \S\ref{sec:lowzrich}.
}
\label{fig:rich}
\end{figure}

With our large cluster sample, we can examine the LF for subsamples of our data of varying richness.  For this test we choose to split the sample into two bins using \bgc~cuts of 300$<$\bgc$\le$500 and \bgc$>$800.  These subsamples are denoted 'poorer' and 'richer' respectively.  The typical error on \bgc~is $\sim$20\%-30\%, so these limits ensure that the bins are as independent as possible, whilst maintaining reasonable numbers in each.  In Fig.~\ref{fig:rich} we show the evolution of the luminous-to-faint ratio, as in Fig.~\ref{fig:evol_gd}, but this time split by richness.  The linear fit from Fig.~\ref{fig:evol_gd} is overplotted as the solid line for reference.  We see a trend in the direction that, at lower redshifts, richer clusters have a lower luminous-to-faint ratio than poorer clusters at the same redshift.  At higher redshifts, z$\sim$0.6, the value of the ratios become indistinguishable within the errors, and may even reverse in the last bin.  

Incompleteness in the cluster catalog is a function of both redshift and richness.  We examine its effects using completeness estimates from \cite{Gladders:2002ui}.  For clusters of \bgc~$\ge$800, the incompleteness is negligible and so the richer bin is unaffected.  However, for the poorest clusters at the highest redshift considered, the incompleteness can reach $\sim$20\%.  To quantify the incompleteness, we integrate over the measured distribution of \bgc~in the poorer subsample at each redshift, applying the completeness corrections, and compare the measured mean \bgc~in each bin with the expected value allowing for incompleteness.  To z$\sim$0.7 the bias in the mean \bgc~in the poorer bin is $\lsim$10\%, but by z$=$0.9 the cluster sample is biased 25\% richer than expected.   This bias may wash out some of the intrinsic difference between richer and poorer clusters at the high redshift end, if the difference seen at lower redshifts still exists there. Lowering the \bgc$>$300 criterion to mitigate the effect of this bias would likely introduce larger systematics, as the false positive rate is expected to significantly increase below this richness. 
  
A potentially more serious selection effect concerns the use of \bgc~cuts.  The clusters are selected by the number of galaxies on their red-sequence within a fixed physical radius.  The magnitude limit adopted for the   \bgc~measurement corresponds to $\sim M^\star+$2 or the 100\% completeness limit, whichever is brighter.  This is faint enough to be affected by the decreasing fraction of fainter red-sequence galaxies at higher redshifts.  If clusters naturally exhibit a monotonic sequence of luminous-to-faint ratios which increases with decreasing richness, and we select clusters based on the total number of galaxies on the red-sequence, then this might impose a limit to the maximum luminous-to-faint ratio we can measure for the poorest clusters.  This occurs since the poorest clusters appear to have a greater deficit of faint red members, and thus such systems with high luminous-to-faint ratios (low fractions of faint galaxies) will be systematically excluded from our sample.  The fact that the two or three highest redshift bins for the poorer clusters show approximately constant luminous-to-faint ratios suggests that we might be seeing such a bias in our sample.
  
Over the redshift range 0.4$\lsim$z$\lsim$0.8, \citet{de-Lucia:2007li}, when splitting their sample by velocity dispersion, found a trend in the opposite direction to that which we see: they suggested that more massive clusters exhibited higher luminous-to-faint ratios than less massive clusters.  We note that the 600 km s$^{-1}$ division they used would correspond to a richness of \bgc$\approx$600 \citep{blindert07a}, which is very close to the dividing line between our richer and poorer clusters.

\subsection{Total cluster luminosity functions}
\label{sec:totlf}
Next we consider luminosity functions for galaxies of all colors.  In order to more fairly compare bluer galaxies with their red-sequence counterparts we apply additional corrections to the former to remove type-dependent star-formation differences so that the \zpri~magnitudes more closely sample the underlying old stellar populations.  Otherwise, blue galaxies temporarily brightened by on-going star-formation would enter our sample and then fade out again at lower redshift as their star-formation rate decreases.  This is akin to deriving a psuedo-stellar mass function, with the luminosity due to star formation removed.  Firstly, we infer a k-correction by comparing the observed galaxy colors with SEDs from \citet{Coleman:1980td}, shifted to the redshift of the composite cluster.  Secondly, we add a simple (model-dependent but small) evolution correction to account for the different average star-formation histories of the red and blue galaxies.  We use a correction term of the form $Q \times z$ where we adopt Q$=$0.9 for the red-sequence galaxies and Q$=$0.6 for bluer galaxies, in the \zpri~band.  The Q$=$0.9 parameterization is an excellent approximation to the luminosity evolution of the z$=$3 burst model we have adopted.  Q$=$0.6 is a reasonable choice for later spectral types (e.g., \citealt{Yee:2005hx}), but the exact choice makes little ($\lsim$0.1 mag) difference to the final differential correction over the redshift range considered here, since $\Delta$Q$<$0.3 and $\Delta$z=0.5.  

\begin{figure*}
\epsscale{1.}
\plotone{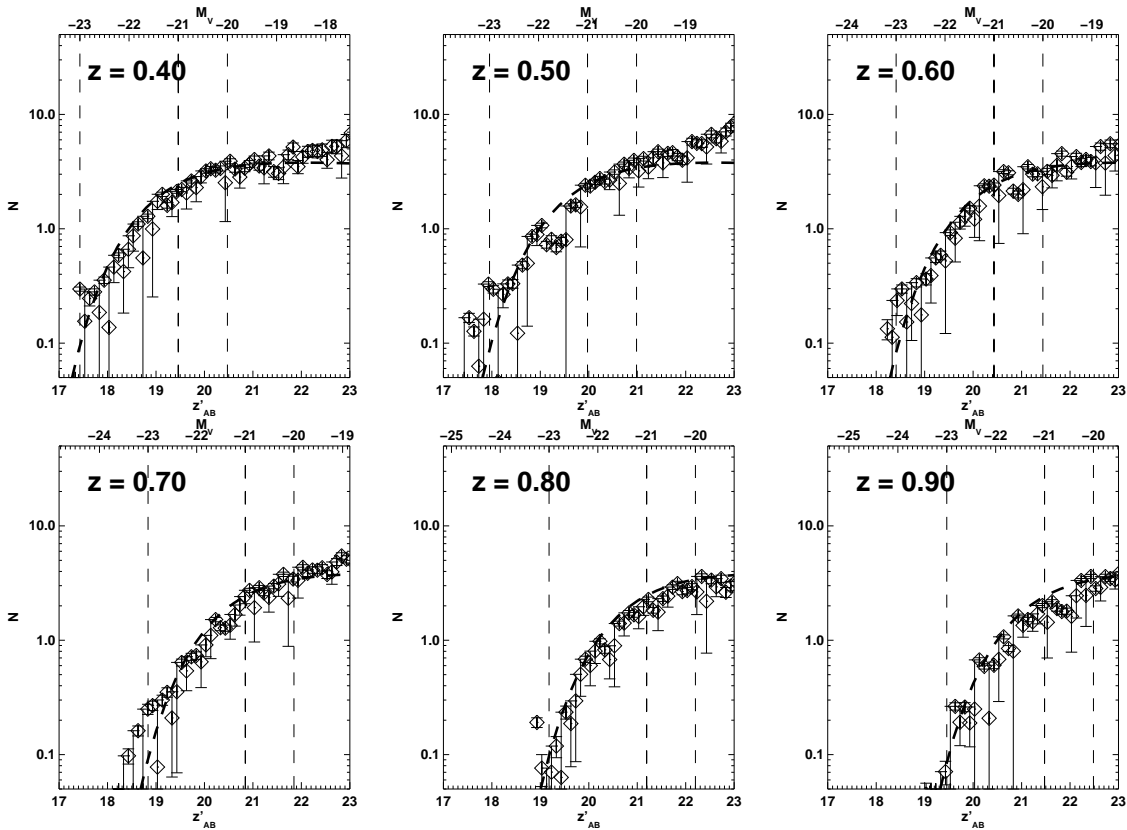}
\caption{As Fig.~\ref{fig:lfs} but for cluster galaxies of all colors. The blue cloud galaxies have been k+e-corrected relative to red-sequence galaxies.  The dashed line is the same passively evolved red-sequence fit as in Fig.~\ref{fig:lfs}.
\label{fig:totlfs}
}
\end{figure*}

In Fig.~\ref{fig:totlfs} we plot composite luminosity functions for cluster galaxies of all colors.  Overlaid is the same curve as shown in Fig.~\ref{fig:lfs}, showing the passively evolved fit to the z$=$0.4 red-sequence LF.  This shows much closer agreement between the evolution expected for red-sequence galaxies and the observed evolution of the total galaxy population of all colors than with the observed evolution of just the red-sequence LF (c.f. data points in Figs.~\ref{fig:lfs} and \ref{fig:totlfs} with dashed line in these figures).  Fig.~\ref{fig:lfs} showed a deficit of faint red galaxies at higher redshift and a slow fall-off in the number of bright red galaxies with increasing redshift.  When galaxies of all colors are considered, the LF (Fig.~\ref{fig:totlfs}) more closely resembles that of the passively evolved z$=$0.4 red-sequence LF.

As an additional step, we examine the differences due to the different stellar mass-to-light ratios, M/L,  of the red and blue galaxies by modelling the blue galaxies as a stellar population with an e-folding timescale of 4 Gyr.  As pointed out by \citet{Bell:2004lb}, this color would correspond approximately to an Sb galaxy locally.  This is a reasonable choice for a typical ``blue cloud" cluster member \citep{Loh:2007be}.  We use the difference in M/L ratio between this model and that of the z$=$3 burst to measure  differential corrections for the bluer galaxies relative to the red-sequence.  Applying this additional correction gives a stellar mass function for the total galaxies which still more closely resembles the expected evolution for the passively evolved red-sequence stellar mass function than does the observed evolution of the red-sequence.  We note that our results are unchanged if instead we simply use the difference between the 4 Gyr e-folding model and the z$=$3 burst model to infer the differential k-, evolution, and M/L corrections, and use this as a check of our semi-empirical corrections.  Given the uncertainties inherent in inferring star-formation histories for blue galaxies from a single color and modelling the entire blue cloud as a single, simple stellar population, we do not pursue this any further here, but regard these results as illustrative.

If we omit the corrections to the bluer galaxies just described and consider simply the uncorrected \zpri-band LF, then the agreement between the observed LF would even more closely trace the curve expected for the passively-evolved red-sequence shown in Fig~\ref{fig:totlfs}.  Thus, we have adopted a more conservative approach by applying these corrections instead of using the \zpri-band LF directly.  Detailed modelling of the blue populations within these clusters is beyond the scope of this current paper, but will be explored in future work.  However, this is very suggestive that the simplest explanation for the evolution of the red-sequence LF is that red-sequence galaxies can be built up from the termination of star-formation in blue cloud galaxies.  

\section{Discussion}
We have constructed a pseudo-mass selected composite cluster sample from a large number of rich galaxy clusters out to z$\sim$1.  For the first time, we have sufficient numbers of cluster members to study in detail the evolution of the luminosity function in an homogeneously selected cluster sample.

\subsection{Faint-end evolution}

At the faint end ($M_V\gsim$-21), the red-sequence luminosity function in clusters declines with increasing redshift.  This is most clearly seen in the evolution of the luminous-to-faint ratio (Fig.~\ref{fig:evol_gd}). Such a decline is consistent with the `down-sizing' picture, in which star-formation proceeds from the most massive to least massive galaxies as the universe ages.  The red-sequence traces the history of star-formation via the ``red and dead" remnants of once-actively star-forming galaxies.  

\label{sec:lowz}

In order to extend the time baseline covered by our RCS-1 sample, we compare with a low redshift cluster sample, taking data from \citet{barkhouse07a} who studied the red-sequence luminosity functions of a sample of X-ray luminous Abell clusters in the redshift range 0.04$<$z$<$0.20.  
Their method for selecting red-sequence galaxies was very similar to ours, except using $(B-R)$ imaging (which is very close to rest-frame $(R-z^\prime)$ at z$\sim$0.5).  We simply convert their $R$-band photometry into the rest frame $V$-band using the same stellar population models described in \S\ref{sec:rslf} and converting their assumed $h$=0.5 to $h$=0.7.  The luminous-to-faint ratio for this sample is shown as the open diamond in Fig.~\ref{fig:evol_gd}.  This low-redshift value is consistent with the extrapolation of the linear fit to our higher redshift data.  This implies that, within the luminosity range we are probing, down-sizing is still on-going from z$\sim$0.9 down to the present day.  

\subsection{Cluster selection}
In a hierarchical universe, clusters observed at some given epoch will have less massive progenitors at higher redshift.  Therefore, it is pertinent to ask how the clusters at the low redshift end of our sample relate to those at the high redshift end.  To do this, we use the results of \citet{van-den-Bosch:2002zq} who used N-body simulations to study the merger histories of cold dark matter haloes.  From his fig.~3, it can be seen that the main progenitor of a halo of mass typical for the clusters being studied here would grow in mass on average by a factor of $\lsim$2 between a redshift of 0.8 and 0.4. Our mass proxy for selecting our sample is \bgc.  Assuming the relation between \bgc~and mass does not evolve between z$=$0.4 and z$=$0.9, this would correspond to a change in the \bgc~limit from 500 at the low redshift end to 350 at the high redshift end.  We have tried applying an evolving limit following a similar prescription and found that it does not affect our main results concerning the evolution of the luminous-to-faint ratio in clusters.  \citet{Gladders:2007us} found from a self-calibration technique used to derive cosmological parameters that the best-fit mass--richness relation was compatible with no redshift evolution, but with a large uncertainty.  Work is on-going to establish whether the assumption that the \bgc--mass relation does not evolve with redshift is valid.  This is an observationally challenging project, but early results \citep{hicks07, felipe07} suggest that the higher redshift relation, z$\gsim$0.7-1.1, appears consistent with the lower redshift calibration adopted here. If anything, the relation may evolve slightly in the direction that causes a given \bgc~to represent a less massive cluster at higher redshift, which would mean that the non-evolving \bgc~limit used here may naturally account for evolution in the average cluster mass through hierarchical growth.

The most serious aspect of the selection which may potentially influence the results is that of imposing a red-sequence richness cut.  \bgc~is calculated from the number of red-sequence members brighter than M$^\star$+2 (or the 100\% completeness limit, whichever is brighter) within a 0.5 $h^{-1}_{50}$ Mpc radius aperture.  The coefficients associated with this measure are designed to make the value insensitive to the choice of counting radius or magnitude range under the assumption of a universal luminosity function.  However, we have shown that the faint end of the luminosity function for red-sequence galaxies both evolves relative to the bright end with redshift and depends on cluster richness.  Below z$\sim$0.7, for the average cluster in this study, the down-sizing effect has little impact on our \bgc~measurements as any effect occurs below the magnitude limit used for \bgc.  Indeed, our red-sequence richness selection should preferentially pick systems least affected by down-sizing and so our results concerning the faint-end deficit of red-sequence galaxies should be regarded as a lower limit.  The good agreement of our luminous-to-faint ratios with other work (despite using slightly different magnitude ranges) suggests that our results are not seriously affected, though.  Furthermore, the fact that mass estimates of individual clusters agree well with the expected masses based on \bgc~at high redshift \citep{gilbank:07a,felipe07} and that cosmological constraints based on cluster abundances \citep{Gladders:2007us} agree with concordance values, suggest that this effect is likely not of serious concern for most of our studies.  However, the cluster-dependent effects of down-sizing must be considered, especially for the lowest mass systems (Fig.~\ref{fig:rich}) when studying galaxy evolution in our clusters.  Accounting for these effects suggests a higher order correction which may improve our mass estimates based on \bgc.

Projection of unrelated structure along the line of sight must also be considered. It is expected from simulations  \citep{Gladders:2002ui} and also measured from spectroscopy \citep{blindert07a,gilbank:07a}, that $\sim$5-10\% of red-sequence clusters in our study will have another cluster which is a significant fraction of the richness of the parent cluster projected onto their red-sequence.  However, in order to be close enough in redshift space that the red-sequences cannot be distinguished by their $(R-z^\prime)$ colors, the clusters would typically be closer than $\Delta$z$=$0.1, the size of our redshift bins.  So, the effect of projections is to some extent mitigated by our bin size, as the clusters are likely to be still placed in the same bin.  The main problem is that clusters will be slightly undercounted, i.e., two clusters are placed in the same bin, but only counted as one.  Projections where the clusters should be placed in different bins will just contribute a certain amount of cross-talk between bins, blurring the difference between different redshifts.  However, the magnitude of this effect is likely to be $\lsim$5\% from the argument above. Thus, the net effect of projection will mostly be to add noise to the RSLFs, and not to artificially produce any of the trends we find.  The richness dependence of the luminous-to-faint ratio means that a slight bias may be introduced in that the projection of two poorer clusters which appear as one richer cluster should have a slightly higher luminous-to-faint ratio than expected for a single richer cluster.  Again, the fraction of systems so affected is likely to be small.

\citet{Cohn:2007cr} recently used a red-sequence selection to find clusters within the Millennium simulation.  Their algorithm differs in detail from ours and there are still numerous important unresolved issues such as the density dependent colours of galaxies within the semi-analytic models used, but they estimate contamination rates of $\sim$10\% at z$=$0.4 and $\sim$20\% at z$=$0.9.  

\subsection{Total luminosity functions}
Fig.~\ref{fig:lfs} shows that, moving from high redshift towards the current epoch, the faint end of the red-sequence luminosity function (or, equivalently the stellar mass function) becomes increasingly populated.  Fig.~\ref{fig:totlfs} shows that, after attempting to correct bluer galaxies' luminosities to stellar masses (which would otherwise allow those temporarily brightened by on-going star-formation to artificially enter our sample), the stellar mass function of {\it all} galaxies in an average cluster appears relatively constant.  Thus, the build-up of red galaxies appears to be accompanied by a decline in blue galaxies in clusters.  This suggests that a simple explanation for the build up of the red-sequence is the conversion of blue to red galaxies due to the termination of star-formation.

The study of the evolution of the overall normalization of the luminosity/stellar mass function is complicated by the fact that our cluster mass estimator is based on cluster richness.  Thus, evolution in number density of galaxies within the cluster is degenerate with evolution in cluster mass.  However, we can use the assumption that the average cluster in the high redshift bins evolves into the average cluster in the lower redshift bins to justify the above argument.  We showed in the previous section that, within the uncertainties of our sample selection, this is likely not a bad assumption.

Hence, with the above caveats, we can say that not only do we see an increasing population of blue galaxies toward the faint end of the LF at higher redshift, but we see an increase of blue galaxies at the bright end also (c.f., Fig.~\ref{fig:lfs} and Fig.~\ref{fig:totlfs}).  Recent observations have found signs of mergers between bright, red galaxies in a handful of high redshift clusters \citep{2004astro.ph..8165F,Tran:2005xc,Mei:2006mu} and such a ``dry merging'' mechanism has been proposed for explaining the growth of the bright end of the red-sequence.  While we cannot place limits on the incidences of such mergers in our sample (due to the degeneracy between mass and number density), our finding of an increase in the number of bright blue galaxies in clusters at these redshifts suggests that the termination of star-formation may also be a significant mechanism.  We note that although the bright end of cluster luminosity functions locally are dominated by red-sequence galaxies, spectroscopically confirmed blue cluster members as bright as the brightest cluster galaxies are seen to exist in clusters at z$\sim$1 \citep{Mei:2006mu}.  In addition, in a sample of z$\sim$1 clusters, \citet{2004astro.ph..8165F} suggested that the brightest cluster galaxies had considerable evolution ahead of them.  In 3 of their 6 clusters, the BCGs had morphologies of S0 or later, and in at least one cluster, the BCG seemed to comprise a pair of galaxies, close enough to potentially merge.  This is support for our finding that the red-sequence LF might be produced simply by conversion of star-forming galaxies to passively evolving galaxies.  

A more thorough treatment of this problem requires the use of an infall model and more detailed modelling of the stellar populations, and such work will be presented in a future paper.

\subsection{Richness dependence}
\label{sec:lowzrich}
Before looking at richness dependence within our own cluster sample, we can compare our results with the COMBO-17 field sample of \citep{Bell:2004lb}.  In their fig.~3, \cite{Bell:2004lb} plot a reference Schechter function with faint end slope $\alpha=$-0.6.  Their actual data begin to fall off faster than this -0.6 line, toward higher redshift.  In our cluster data, the faint end red-sequence slope never turns over as quickly as this -0.6 line.  i.e., the field will always show a greater deficit of faint galaxies in the RSLF relative to our cluster data.  This suggests that star-formation ended later in faint galaxies in the field than in clusters.  This can be viewed as an extension of the down-sizing phenomenon: not only does star-formation progress from more massive to less massive galaxies as the universe ages, but it also progresses from more massive to less massive clusters.

For the cluster subsamples split by richness (Fig.~\ref{fig:rich}), the richer sample (\bgc $>$ 800) traces the best-fit linear relation (Fig.~\ref{fig:evol_gd}) to the whole sample (\bgc $>$ 500), only with larger error bars due to fewer galaxies, again demonstrating that the richness cut chosen at this level does not affect the results.  A low redshift comparison point from a rich (\bgc$>$1000) subsample of the \citet{barkhouse07a} data is also included and found to fit with the linear relation extrapolated from our higher redshift sample. The dashed line shows a power law fit of the form $\propto (1+z)^\beta$.  We find a best fit $\beta=(1.90\pm0.35)$, which only modestly differs from the linear fit over the redshift range probed by RCS, but gives better agreement with the low redshift point from the \citealt{barkhouse07a} sample.  Recently, \citet{Stott:2007wc} parameterized the evolution in the luminous-to-faint ratio in this way, finding a value of $\beta=(2.5\pm0.5)$, which agrees with our value.  They used an X-ray luminous sample (of 10 clusters at z$\sim$0.1 and 10 at z$\sim$0.5 plus several additional clusters), which should be most comparable to our rich subsample here.  However, they probe to a fainter magnitude limit (similar to the \citealt{De-Lucia:2006sa} and \citealt{Tanaka:2005mk}  depths) and so it is not obvious that the form of the redshift evolution should be the same as for our sample.  We note that the trend they found is in good qualitative agreement with our work and reasonable quantitative agreement.

At the low redshift end of our RCS sample, we find that the poorer clusters have systematically higher luminous-to-faint ratios than the rich clusters, meaning that their faint end RSLF is falling off more rapidly than that of the richer clusters.  This means that poorer clusters more closely resemble the field than do rich clusters.  Again, none of our poorer sample's RSLF falls off as quickly as the upper limit to the field value of -0.6 found by \citet{Bell:2004lb} and thus, poorer clusters have RSLFs which appear intermediate between rich clusters and the field.  This is in the opposite direction to the result quoted by \citet{De-Lucia:2006sa} in the redshift range 0.4$\lsim$z$\lsim$0.8 from their sample of $\sim$10 clusters.  We suggest that the most likely cause of this discrepancy is the choice of clusters.  As noted by \citet{De-Lucia:2006sa}, there is significant variation in the RSLFs from cluster to cluster, and our larger sample is more likely to be representative of the average cluster population.  Toward higher redshift, we cannot measure a significant difference between richer and poorer, given the size of our error bars, except in the highest redshift (z$\sim$0.9) bin where the trend appears to reverse.  However, the last bin is the one which is most likely to be affected by incompleteness (both in the cluster sample and in the photometry), so we choose to be conservative and disregard this last bin.  Additionally, incompleteness effects at high redshift act to preferentially remove poorer clusters, causing the poorer bin to shift to systematically higher richnesses, lowering the significance of the difference between the two subsamples.  We have overlaid a best-fit linear relation in Fig.~\ref{fig:rich} to guide the eye, which shows that a couple of points are discrepant with such a fit at the $>$1-$\sigma$ level.  This suggests that our error bars may be slightly underestimated.  We note that the richer (\bgc $>$800) points in Fig.~\ref{fig:rich} could be replaced with the main sample (\bgc $>$500) plotted in Fig.~\ref{fig:evol_gd} to produce a sample with smaller error bars which would still produce an average composite cluster which is significantly richer than the poorer (300$<$\bgc$\le$500) sample, but with more cross-contamination at the boundary of the richness bins, where the median \bgc~error on an individual cluster is $\Delta$\bgc$\sim$200.  This would still produce a significant difference between the richer and poorer clusters at 0.4$\lsim$z$\lsim$0.5, even after increasing the size of the poorer cluster error bars to be consistent with the best fit straight line. 

The results presented imply that the faint end of the red-sequence was built up first in rich clusters, then in poorer clusters and finally in the field.

\section{Conclusions}
We have studied the properties of red-sequence galaxies in a well-defined statistical sample of galaxy clusters over the redshift range 0.35$<$z$\le$0.95.  Each redshift bin of width $\Delta$z$=$0.1 contains $\approx$50 clusters and $\approx$5000 red-sequence galaxies.  Our main results are:

1) The faint end of the red-sequence, as measured by the ratio of luminous-to-faint galaxies, declines with increasing redshift.  This implies that star formation has not yet ended in the faintest cluster galaxies at high redshift.  The red-sequence is built up at the faint end as star-formation proceeds to progressively less luminous (less massive) galaxies, consistent with the down-sizing scenario \citep{Cowie:1996xw}.

2) The turnover of the faint end of the RSLF is dependent on the cluster richness (mass) in the sense that for more massive clusters, the deficit of faint red-sequence galaxies is less than that for less massive clusters.  This is an indication that star formation ended earlier for faint galaxies in richer clusters than in poorer clusters. 

3) The decline in the faint end of the red-sequence toward higher redshift is accompanied by an increase in the total (i.e., including blue galaxies) cluster LF.  This suggests that the build up of faint, red galaxies may be driven largely by the termination of star-formation in low mass galaxies.  A similar increase of blue galaxies is also seen at the brighter end of the LF, suggesting that (at least some of) the build up of high mass early-type galaxies may also be attributed to the termination of star-formation.

Future work will add the $B$- and $V$-band imaging of RCS-1 fields \citep{Hsieh:2005fq} and the accompanying photometric redshift catalog to examine the luminosity functions of RCS clusters.  The $\sim$1000 square degree next generation survey, RCS-2 \citep{Yee:2007qb}, will provide an order of magnitude larger sample to improve upon the statistics of the current work.

\acknowledgements
We thank Bob Abraham, Mike Balogh and Adam Muzzin for helpful discussions.\\  

The RCS project is supported by grants to H.K.C.Y. from the Canada Research Chair Program,  the Natural Sciences and Engineering Research Council of Canada (NSERC) and the University of Toronto.  E.E. acknowledges NSF grant  AST-0206154.  M.D.G. acknowledges partial support for this work provided by NASA through Hubble Fellowship grant HF-01184.01 awarded by the Space Telescope Science Institute, which is operated by the Association of Universities for Research in Astronomy, Inc., for NASA, under contract NAS 5-26555.  L.F.B. acknowledges the support of the FONDAP center for Astrophysics and CONICYT under proyecto  FONDECYT 1040423.  W.B. acknowledges support from NASA LTSA award NAG5-11415, NASA Chandra X-ray Center archival research grant AR7-8015B, and a University of Illinois seed funding award to the Dark Energy Survey.

\label{lastpage}

\begin{thebibliography}{54}
\expandafter\ifx\csname natexlab\endcsname\relax\def\natexlab#1{#1}\fi

\bibitem[{{Baldry} {et~al.}(2006){Baldry}, {Balogh}, {Bower}, {Glazebrook},
  {Nichol}, {Bamford}, \& {Budavari}}]{Baldry:2006bq}
{Baldry}, I.~K., {Balogh}, M.~L., {Bower}, R.~G., {Glazebrook}, K., {Nichol},
  R.~C., {Bamford}, S.~P., \& {Budavari}, T. 2006, \mnras, 373, 469

\bibitem[{{Barkhouse} {et~al.}(2007){Barkhouse}, {Yee}, \&
  {Lopez-Cruz}}]{barkhouse07a}
{Barkhouse}, W., {Yee}, H.~K.~C., \& {Lopez-Cruz}, O. 2007, \apj, in press, astro-ph/0709.0983

\bibitem[{{Barrientos} {et~al.}(2007){Barrientos}, {Gilbank}, {Gladders},
  {Yee}, {Infante}, {Ellingson}, {Hall}, \& {Hertling}}]{felipe07}
{Barrientos}, L.~F., {Gilbank}, D.~G., {Gladders}, M.~D., {Yee}, H.~K.~C.,
  {Infante}, L., {Ellingson}, E., {Hall}, P.~B., \& {Hertling}, G. 2007, in
  prep

\bibitem[{{Baugh} {et~al.}(1996){Baugh}, {Cole}, \& {Frenk}}]{Baugh:1996vs}
{Baugh}, C.~M., {Cole}, S., \& {Frenk}, C.~S. 1996, \mnras, 283, 1361

\bibitem[{{Bell} {et~al.}(2004){Bell}, {Wolf}, {Meisenheimer}, {Rix}, {Borch},
  {Dye}, {Kleinheinrich}, {Wisotzki}, \& {McIntosh}}]{Bell:2004lb}
{Bell}, E.~F., et al. 2004,
  \apj, 608, 752

\bibitem[{{Blindert} {et~al.}(2007){Blindert}, {Yee}, {Gladders}, {Ellingson},
  {Gilbank}, {Barrientos}, \& {Golding}}]{blindert07a}
{Blindert}, K., {Yee}, H.~K.~C., {Gladders}, M.~D., {Ellingson}, E., {Gilbank},
  D.~G., {Barrientos}, L.~F., \& {Golding}, J. 2007, \apjs, submitted

\bibitem[{{Bower} {et~al.}(1992){Bower}, {Lucey}, \& {Ellis}}]{bow92}
{Bower}, R.~G., {Lucey}, J.~R., \& {Ellis}, R.~S. 1992, \mnras, 254, 601+

\bibitem[{{Bruzual} \& {Charlot}(2003)}]{Bruzual:2003de}
{Bruzual}, G. \& {Charlot}, S. 2003, \mnras, 344, 1000

\bibitem[{{Bundy} {et~al.}(2005){Bundy}, {Ellis}, \&
  {Conselice}}]{Bundy:2005bc}
{Bundy}, K., {Ellis}, R.~S., \& {Conselice}, C.~J. 2005, \apj, 625, 621

\bibitem[{{Cimatti} {et~al.}(2006){Cimatti}, {Daddi}, \&
  {Renzini}}]{Cimatti:2006vz}
{Cimatti}, A., {Daddi}, E., \& {Renzini}, A. 2006, \aap, 453, L29

\bibitem[{{Cohn} {et~al.}(2007){Cohn}, {Evrard}, {White}, {Croton}, \&
  {Ellingson}}]{Cohn:2007cr}
{Cohn}, J.~D., {Evrard}, A.~E., {White}, M., {Croton}, D., \& {Ellingson}, E.
  2007, astro-ph/0706.0211

\bibitem[{{Coleman} {et~al.}(1980){Coleman}, {Wu}, \&
  {Weedman}}]{Coleman:1980td}
{Coleman}, G.~D., {Wu}, C.-C., \& {Weedman}, D.~W. 1980, \apjs, 43, 393

\bibitem[{{Colless}(1989)}]{colless89}
{Colless}, M. 1989, \mnras, 237, 799

\bibitem[{{Cowie} {et~al.}(1996){Cowie}, {Songaila}, {Hu}, \&
  {Cohen}}]{Cowie:1996xw}
{Cowie}, L.~L., {Songaila}, A., {Hu}, E.~M., \& {Cohen}, J.~G. 1996, \aj, 112,
  839

\bibitem[{{De Lucia} {et~al.}(2004){De Lucia}, {Poggianti},
  {Arag{\'o}n-Salamanca}, {Clowe}, {Halliday}, {Jablonka}, {Milvang-Jensen},
  {Pell{\'o}}, {Poirier}, {Rudnick}, {Saglia}, {Simard}, \&
  {White}}]{De-Lucia:2004xa}
{De Lucia}, G., et al. 2004, \apjl, 610, L77

\bibitem[{{de Lucia} {et~al.}(2007){de Lucia}, {Poggianti},
  {Arag{\'o}n-Salamanca}, {White}, {Zaritsky}, {Clowe}, {Halliday}, {Jablonka},
  {von der Linden}, {Milvang-Jensen}, {Pell{\'o}}, {Rudnick}, {Saglia}, \&
  {Simard}}]{de-Lucia:2007li}
{De Lucia}, G., et al. 2007, \mnras, 374, 809

\bibitem[{{De Lucia} {et~al.}(2006){De Lucia}, {Springel}, {White}, {Croton},
  \& {Kauffmann}}]{De-Lucia:2006sa}
{De Lucia}, G., {Springel}, V., {White}, S.~D.~M., {Croton}, D., \&
  {Kauffmann}, G. 2006, \mnras, 366, 499

\bibitem[{{Eggen} {et~al.}(1962){Eggen}, {Lynden-Bell}, \&
  {Sandage}}]{Eggen:1962yu}
{Eggen}, O.~J., {Lynden-Bell}, D., \& {Sandage}, A.~R. 1962, \apj, 136, 748

\bibitem[{{Faber} {et~al.}(2005){Faber}, {Willmer}, {Wolf}, {Koo}, {Weiner},
  {Newman}, {Im}, {Coil}, {Conroy}, {Cooper}, {Davis}, {Finkbeiner}, {Gerke},
  {Gebhardt}, {Groth}, {Guhathakurta}, {Harker}, {Kaiser}, {Kassin},
  {Kleinheinrich}, {Konidaris}, {Lin}, {Luppino}, {Madgwick}, {Noeske},
  {Phillips}, {Sarajedini}, {Simard}, {Szalay}, {Vogt}, \&
  {Yan}}]{Faber:2005yw}
{Faber}, S.~M., et al. 2005, astro-ph/0506044

\bibitem[{{Ford} {et~al.}(2004){Ford}, {Postman}, {Blakeslee}, {Demarco},
  {Jee}, {Rosati}, {Holden}, {Homeier}, {Illingworth}, \&
  {White}}]{2004astro.ph..8165F}
{Ford}, H., et al. 2004, in Astrophysics and Space Science Library, Vol. 319, Penetrating
  Bars Through Masks of Cosmic Dust, ed. D.~L. {Block}, I.~{Puerari}, K.~C.
  {Freeman}, R.~{Groess}, \& E.~K. {Block}, 459--+

\bibitem[{{Gilbank} {et~al.}(2007){Gilbank}, {Yee}, {Ellingson}, {Gladders},
  Barrientos, \& {Blindert}}]{gilbank:07a}
{Gilbank}, D.~G., {Yee}, H.~K.~C., {Ellingson}, E., {Gladders}, M.~D.,
  Barrientos, L.~F., \& {Blindert}, K. 2007, \aj, 134, 282

\bibitem[{{Gladders}(2002)}]{Gladders:2002ui}
{Gladders}, M.~D. 2002, Ph.D.~Thesis, University of Toronto

\bibitem[{{Gladders} \& {Yee}(2000)}]{gy00}
{Gladders}, M.~D. \& {Yee}, H.~K.~C. 2000, \aj, 120, 2148

\bibitem[{{Gladders} \& {Yee}(2005)}]{Gladders:2005oi}
---. 2005, \apjs, 157, 1

\bibitem[{{Gladders} {et~al.}(2007){Gladders}, {Yee}, {Majumdar}, {Barrientos},
  {Hoekstra}, {Hall}, \& {Infante}}]{Gladders:2007us}
{Gladders}, M.~D., {Yee}, H.~K.~C., {Majumdar}, S., {Barrientos}, L.~F.,
  {Hoekstra}, H., {Hall}, P.~B., \& {Infante}, L. 2007, \apj, 655, 128

\bibitem[{{Hicks} {et~al.}(2006){Hicks}, {Ellingson}, {Hoekstra}, \&
  {Yee}}]{Hicks:2006ap}
{Hicks}, A.~K., {Ellingson}, E., {Hoekstra}, H., \& {Yee}, H.~K.~C. 2006, \apj,
  652, 232

\bibitem[{{Hicks} {et~al.}(2007){Hicks}, {Ellingson}, {Yee}, {Gladders},
  {Hoekstra}, {Gilbank}, {Bautz}, {Cain}, \& {Garmire}}]{hicks07}
{Hicks}, A.~K., et al. 2007, in prep

\bibitem[{{Hsieh} {et~al.}(2005){Hsieh}, {Yee}, {Lin}, \&
  {Gladders}}]{Hsieh:2005fq}
{Hsieh}, B.~C., {Yee}, H.~K.~C., {Lin}, H., \& {Gladders}, M.~D. 2005, \apjs,
  158, 161

\bibitem[{{Juneau} {et~al.}(2005){Juneau}, {Glazebrook}, {Crampton},
  {McCarthy}, {Savaglio}, {Abraham}, {Carlberg}, {Chen}, {Le Borgne}, {Marzke},
  {Roth}, {J{\o}rgensen}, {Hook}, \& {Murowinski}}]{Juneau:2005ft}
{Juneau}, S., et al. 2005, \apjl, 619, L135

\bibitem[{{Kauffmann} \& {Charlot}(1998)}]{Kauffmann:1998mh}
{Kauffmann}, G. \& {Charlot}, S. 1998, \mnras, 294, 705

\bibitem[{{Loh} {et~al.}(2007){Loh}, {Ellingson}, {Yee}, {Gilbank}, {Gladders},
  \& {Barrientos}}]{Loh:2007be}
{Loh}, Y.-S., {Ellingson}, E., {Yee}, H.~K.~C., {Gilbank}, D.~G., {Gladders},
  M.~D., \& {Barrientos}, L.~F. 2007, ApJ, submitted

\bibitem[{{Longair} \& {Seldner}(1979)}]{ls79}
{Longair}, M.~S. \& {Seldner}, M. 1979, \mnras, 189, 433

\bibitem[{{Lumsden} {et~al.}(1992){Lumsden}, {Nichol}, {Collins}, \&
  {Guzzo}}]{edcc}
{Lumsden}, S.~L., {Nichol}, R.~C., {Collins}, C.~A., \& {Guzzo}, L. 1992,
  \mnras, 258, 1

\bibitem[{{McNamara} \& {O'Connell}(1992)}]{McNamara:1992wj}
{McNamara}, B.~R. \& {O'Connell}, R.~W. 1992, \apj, 393, 579

\bibitem[{{Mei} {et~al.}(2006){Mei}, {Blakeslee}, {Stanford}, {Holden},
  {Rosati}, {Strazzullo}, {Homeier}, {Postman}, {Franx}, {Rettura}, {Ford},
  {Illingworth}, {Ettori}, {Bouwens}, {Demarco}, {Martel}, {Clampin}, {Hartig},
  {Eisenhardt}, {Ardila}, {Bartko}, {Ben{\'{\i}}tez}, {Bradley}, {Broadhurst},
  {Brown}, {Burrows}, {Cheng}, {Cross}, {Feldman}, {Golimowski}, {Goto},
  {Gronwall}, {Infante}, {Kimble}, {Krist}, {Lesser}, {Menanteau}, {Meurer},
  {Miley}, {Motta}, {Sirianni}, {Sparks}, {Tran}, {Tsvetanov}, {White}, \&
  {Zheng}}]{Mei:2006mu}
{Mei}, S., et al. 2006, \apj, 639, 81

\bibitem[{{Nelan} {et~al.}(2005){Nelan}, {Smith}, {Hudson}, {Wegner}, {Lucey},
  {Moore}, {Quinney}, \& {Suntzeff}}]{Nelan:2005ml}
{Nelan}, J.~E., {Smith}, R.~J., {Hudson}, M.~J., {Wegner}, G.~A., {Lucey},
  J.~R., {Moore}, S.~A.~W., {Quinney}, S.~J., \& {Suntzeff}, N.~B. 2005, \apj,
  632, 137

\bibitem[{{Partridge} \& {Peebles}(1967)}]{Partridge:1967nh}
{Partridge}, R.~B. \& {Peebles}, P.~J.~E. 1967, \apj, 147, 868

\bibitem[{{Sandage} {et~al.}(1970){Sandage}, {Freeman}, \&
  {Stokes}}]{Sandage:1970lj}
{Sandage}, A., {Freeman}, K.~C., \& {Stokes}, N.~R. 1970, \apj, 160, 831

\bibitem[{{Scarlata} {et~al.}(2007){Scarlata}, {Carollo}, {Lilly}, {Feldmann},
  {Kampczyk}, {Renzini}, {Cimatti}, {Halliday}, {Daddi}, {Sargent},
  {Koekemoer}, {Scoville}, {Kneib}, {Leauthaud}, {Massey}, {Rhodes}, {Tasca},
  {Capak}, {McCracken}, {Mobasher}, {Taniguchi}, {Thompson}, {Ajiki}, {Aussel},
  {Murayama}, {Sanders}, {Sasaki}, {Shioya}, \& {Takahashi}}]{Scarlata:2007oz}
{Scarlata}, C., et al. 2007, astro-ph/0701746

\bibitem[{{Schechter}(1976)}]{schechter}
{Schechter}, P. 1976, \apj, 203, 297

\bibitem[{{Schlegel} {et~al.}(1998){Schlegel}, {Finkbeiner}, \&
  {Davis}}]{1998ApJ...500..525S}
{Schlegel}, D.~J., {Finkbeiner}, D.~P., \& {Davis}, M. 1998, \apj, 500, 525

\bibitem[{{Stott} {et~al.}(2007){Stott}, {Smail}, {Edge}, {Ebeling}, {Smith},
  {Kneib}, \& {Pimbblet}}]{Stott:2007wc}
{Stott}, J.~P., {Smail}, I., {Edge}, A.~C., {Ebeling}, H., {Smith}, G.~P.,
  {Kneib}, J.~., \& {Pimbblet}, K.~A. 2007, \apj, 661 95

\bibitem[{{Tanaka} {et~al.}(2005){Tanaka}, {Kodama}, {Arimoto}, {Okamura},
  {Umetsu}, {Shimasaku}, {Tanaka}, \& {Yamada}}]{Tanaka:2005mk}
{Tanaka}, M., {Kodama}, T., {Arimoto}, N., {Okamura}, S., {Umetsu}, K.,
  {Shimasaku}, K., {Tanaka}, I., \& {Yamada}, T. 2005, \mnras, 362, 268

\bibitem[{{Terlevich} {et~al.}(2001){Terlevich}, {Caldwell}, \&
  {Bower}}]{Terlevich:2001bx}
{Terlevich}, A.~I., {Caldwell}, N., \& {Bower}, R.~G. 2001, \mnras, 326, 1547

\bibitem[{{Tran} {et~al.}(2005){Tran}, {van Dokkum}, {Franx}, {Illingworth},
  {Kelson}, \& {Schreiber}}]{Tran:2005xc}
{Tran}, K.-V.~H., {van Dokkum}, P., {Franx}, M., {Illingworth}, G.~D.,
  {Kelson}, D.~D., \& {Schreiber}, N.~M.~F. 2005, \apjl, 627, L25

\bibitem[{{Tukey}(1958)}]{tukey}
{Tukey}, J.~W. 1958, Ann. Math. Stat., 29, 614

\bibitem[{{van den Bosch}(2002)}]{van-den-Bosch:2002zq}
{van den Bosch}, F.~C. 2002, \mnras, 331, 98

\bibitem[{{van Dokkum}(2005)}]{van-Dokkum:2005dl}
{van Dokkum}, P.~G. 2005, \aj, 130, 2647

\bibitem[{{Visvanathan}(1978)}]{visv}
{Visvanathan}, N. 1978, \aap, 67, L17

\bibitem[{{Yee}(1991)}]{1991PASP..103..396Y}
{Yee}, H.~K.~C. 1991, \pasp, 103, 396

\bibitem[{{Yee} \& {Ellingson}(2003)}]{Yee:2003we}
{Yee}, H.~K.~C. \& {Ellingson}, E. 2003, \apj, 585, 215

\bibitem[{{Yee} {et~al.}(2007){Yee}, {Gladders}, {Gilbank}, {Majumdar},
  {Hoekstra}, {Ellingson}, \& {the RCS-2 Collaboration}}]{Yee:2007qb}
{Yee}, H.~K.~C., {Gladders}, M.~D., {Gilbank}, D.~G., {Majumdar}, S.,
  {Hoekstra}, H., {Ellingson}, E., \& {the RCS-2 Collaboration}. 2007, astro-ph/0701839

\bibitem[{{Yee} {et~al.}(2005){Yee}, {Hsieh}, {Lin}, \&
  {Gladders}}]{Yee:2005hx}
{Yee}, H.~K.~C., {Hsieh}, B.~C., {Lin}, H., \& {Gladders}, M.~D. 2005, \apjl,
  629, L77
  
\bibitem[{{Yee} \& {L{\'o}pez-Cruz}(1999)}]{ylc99}
{Yee}, H.~K.~C. \& {L{\'o}pez-Cruz}, O. 1999, \aj, 117, 1985

\end{thebibliography}
\end{document}